%% file: main.tex
\documentclass[sigconf]{acmart} 
\usepackage{multirow}
\usepackage{float}
\usepackage{url}
\usepackage{algorithm}
\usepackage{algpseudocode}
\usepackage{bm}
\usepackage{graphicx}
\usepackage{caption}
\usepackage[skip=0cm,list=true,labelfont=it]{subcaption}

\theoremstyle{definition}

\AtBeginDocument{%
  \providecommand\BibTeX{{%
    \normalfont B\kern-0.5em{\scshape i\kern-0.25em b}\kern-0.8em\TeX}}}

\copyrightyear{2024}
\acmYear{2024}
\setcopyright{acmlicensed}\acmConference[RecSys '24]{18th ACM Conference
on Recommender Systems}{October 14--18, 2024}{Bari, Italy}
\acmBooktitle{18th ACM Conference on Recommender Systems (RecSys '24),
October 14--18, 2024, Bari, Italy}
\acmDOI{10.1145/3640457.3688149}
\acmISBN{979-8-4007-0505-2/24/10}

\begin{document}

\title{Unlocking the Hidden Treasures: Enhancing Recommendations with Unlabeled Data}

\author{Yuhan Zhao}
\affiliation{
  \institution{Harbin Engineering University}
  \institution{Hong Kong Baptist University}
  \city{Harbin}
  \country{China}
}
\email{asa9ao@hrbeu.edu.cn}

\author{Rui Chen}
\authornote{Corresponding authors.}
\affiliation{
  \institution{Harbin Engineering University}
  \city{Harbin}
  \country{China}
}
\email{ruichen@hrbeu.edu.cn}

\author{Qilong Han}
\authornotemark[1]
\affiliation{
  \institution{Harbin Engineering University}
  \city{Harbin}
  \country{China}
}
\email{hanqilong@hrbeu.edu.cn}

\author{Hongtao Song}
\affiliation{
  \institution{Harbin Engineering University}
  \city{Harbin}
  \country{China}
}
\email{songhongtao@hrbeu.edu.cn}

\author{Li Chen}
\affiliation{
  \institution{Hong Kong Baptist University}
  \city{Hong Kong}
  \country{China}
}
\email{lichen@comp.hkbu.edu.hk}

\input{0.abstract}


\maketitle

\input{1.introduction}

\input{3.preliminary}

\input{4.methodology}
\input{5.experiment}

\input{2.related_work}
\input{7.conclusion}
\input{acks}

\balance
\bibliographystyle{ACM-Reference-Format}
\bibliography{paper}

\end{document}

%% file: 0.abstract.tex
\begin{abstract}
Collaborative filtering (CF) stands as a cornerstone in recommender systems, yet effectively leveraging the massive unlabeled data presents a significant challenge. Current research focuses on addressing the challenge of unlabeled data by extracting a subset that closely approximates negative samples. Regrettably, the remaining data are overlooked, failing to fully integrate this valuable information into the construction of user preferences. To address this gap, we introduce a novel positive-neutral-negative (PNN) learning paradigm. PNN introduces a neutral class, encompassing intricate items that are challenging to categorize directly as positive or negative samples. By training a model based on this triple-wise partial ranking, PNN offers a promising solution to learning complex user preferences. Through theoretical analysis, we connect PNN to one-way partial AUC (OPAUC) to validate its efficacy. Implementing the PNN paradigm is, however, technically challenging because: (1) it is difficult to classify unlabeled data into neutral or negative in the absence of supervised signals; (2) there does not exist any loss function that can handle set-level triple-wise ranking relationships. To address these challenges, we propose a semi-supervised learning method coupled with a user-aware attention model for knowledge acquisition and classification refinement. Additionally, a novel loss function with a two-step centroid ranking approach enables handling set-level rankings. Extensive experiments on four real-world datasets demonstrate that, when combined with PNN, a wide range of representative CF models can consistently and significantly boost their performance. Even with a simple matrix factorization, PNN can achieve comparable performance to sophisticated graph neutral networks. Our code is publicly available at \url{https://github.com/Asa9aoTK/PNN-RecBole}.

\end{abstract}

%% file: 1.introduction.tex
\section{INTRODUCTION}
Collaborative filtering(CF), as the most fundamental technique in recommender systems~\cite{MZW21, SK09, JHX23}, aims to capture users’ preferences and make top-K recommendations based on their historical interactions~\cite{HLZ17,WHW19, HDW20}. In real-world scenarios, these interactions are typically in the form of implicit feedback (e.g., clicks or purchases), rather than explicit ratings~\cite{WFH21}. In these scenarios, user interactions act as positive samples, reflecting user preferences. However, the numerous unobserved items, lack explicit user feedback and remain unlabeled, thereby hindering effective utilization.

Most existing works~\cite{RFG12,CZW19,ZCW13,DQH19,CZZ20} make an intuitive assumption that unlabeled data can directly provide negative signals. Yet, recent works reveal a nuanced reality: \textit{unlabeled data and negative samples harbor an inevitable disparity}. For instance, unlabeled data may contain items that users potentially like or items with user uncertainty. 
Acknowledging this phenomenon, current research endeavors to address the challenge of unlabeled data by attempting to extract a subset closely approximating negative samples, thereby bridging this gap. Techniques like negative sampling \cite{DQY20, ZZH22, CLJ22}, data filtering \cite{MZW21, WFH21}, and weight adjustment \cite{LGZ24} are employed for this purpose. Regrettably, under this approach, the remaining data are overlooked, failing to fully integrate this valuable information into the construction of user preferences. This limitation hampers these methods from fully harnessing the potential of unlabeled data. 

In this paper, we pose a fundamental question: \textit{Is it feasible to fully unearth and leverage the information within massive unlabeled data?} An intuitive approach might be to initially follow the existing methods by filtering out some negative samples and treating the remaining items as positive instances. However, such an approach is ill-suited for recommender systems. Recommendations fundamentally deal with ranking problems rather than binary classification. We cannot guarantee that the remaining items exclusively belong to the positive class. Moreover, the challenge intensifies as hard negative samples are difficult to distinguish from positive samples. Misclassifying hard negative samples as positive instances can severely compromise effectiveness.

Our answer is a novel generic positive-neutral-negative (PNN) learning paradigm. Given user $u \in U$, different from categorizing all items into either positive $\Sigma_{pos}^u$ or negative $\Sigma_{neg}^u$, our insight is to introduce a third ``neutral'' class $\Sigma_{neu}^u$, which encompasses a plethora of intricate items that are challenging to directly categorize as either positive or negative samples. In what follows, we slightly abuse the notation to let $\Sigma_{*}^u$ also denote the set of items with class label $*$. With this new class, our goal is to train a model based on partial rankings $\Sigma_{pos}^u >_u \Sigma_{neu}^u >_u \Sigma_{neg}^u$(i.e., $u$ prefers items $\Sigma_{pos}^u$ over $\Sigma_{neu}^u$ over $\Sigma_{neg}^u$). This way allows complex and challenging unlabeled data to intuitively participate in user preference construction. Furthermore, we establish a theoretical connection between PNN and one-way partial AUC (OPAUC) to validate PNN's efficacy in recommendation performance.

While the conceptual framework of the PNN paradigm effectively addresses the aforementioned issue, providing a concrete implementation is technically challenging for at least two reasons: 
\begin{itemize}
    \item \textbf{Sparse supervised signals.} Reliable signals primarily stem from user interactions, which tend to be sparse compared to the pool of unlabeled data. This scarcity impedes the effective identification of neutral and negative items within massive unlabeled data.
    \item \textbf{Constrained positive-negative loss functions.} Traditional loss functions categorize all items into either positive or negative classes, failing to accommodate the nuanced nature of neutral items. Even if these latent neutral items are unearthed, they are inevitably pigeonholed into positive or negative classifications.
\end{itemize}

To address the former challenge, inspired by pseudo-labelling techniques~\cite{RDR21, WWS22}, we design a novel semi-supervised learning method to leverage the BPR loss to acquire sufficient prior knowledge. This knowledge enables us to discern intricate signals concealed within unlabeled data. Then, we propose a user-aware attention model to indicate the readiness to classify unlabeled data. This model uses classification performance on observed items to decide how to change semi-supervised learning to our loss function. For the second problem, we propose a novel loss function that deals with set-level ranking relationships involving three classes. Though preference ranking may not satisfy the transitive property~\cite{TA69, EI12, MG17}, theoretically we can still \emph{losslessly} decompose a triple-wise ranking into three pairwise rankings. Since $\Sigma_{pos}^u >_u \Sigma_{neg}^u$ can be learned from the BPR loss during the knowledge acquisition phase, we introduce a novel loss function to tackle $\Sigma_{pos}^u >_u \Sigma_{neu}^u$ and $\Sigma_{neu}^u >_u \Sigma_{neg}^u$. To support set-level rankings, we propose an innovative two-step centroid ranking approach. In the first step, we advocate representing each set of items within the three classes by its centroid, thereby transforming intricate set-level relationships into more manageable item-level counterparts. Subsequently, we propose a clamp mechanism meticulously to maintain adaptive margins between distinct classes. This mechanism guarantees that every item in a class ranks higher or lower than any item in another class. We summarize our main contributions as follows:
\begin{itemize}
    \item We introduce the novel PNN learning paradigm, a pioneering approach to fully exploit the wealth of information within massive unlabeled data in CF. By incorporating a third neutral class and leveraging set-level ranking relationships, PNN addresses the complexity of unlabeled data. Furthermore, through mathematical analysis, we establish the relationship between PNN and OPAUC, demonstrating PNN's ability to optimize various indicators of the recommendation system.
    
    \item We propose a concrete implementation of the PNN learning paradigm which can be seamlessly integrated with multiple mainstream CF models. It features a semi-supervised learning method with a user-aware attention model to reliably classify unlabeled data and a two-step centroid ranking approach to accommodate set-level rankings.  
    
    \item We perform extensive experiments on four public datasets to demonstrate that, when combined with PNN, a wide variety of mainstream CF models can consistently and substantially boost their performance, confirming the value of PNN.
\end{itemize}

%% file: 3.preliminary.tex
\section{PNN MEETS OPAUC }
In this section, we embark on a review of traditional methods and shed light on the PNN paradigm, leveraging the OPAUC metric.

The goal of CF is to learn the model's parameters $\Theta$ to recommend the top-ranked items. In the implicit feedback setting, for a user $u$, $\mathcal{I}_u^{+}$ denotes the positive class constituting by observed items while the remainder, $\mathcal{I}_u^{un} = \mathcal{I} \setminus \mathcal{I}_u^{+}$, represents unlabeled data. We demarcate true negative items as $\mathcal{I}_u^{-}$. Traditional loss functions can be formulated as follows:
\begin{equation}
\min_\Theta \sum_{u \in U} \sum_{i \in \mathcal{I}_u^{+}} E_{j \sim P_{ns}(j \mid \Theta )}[\mathcal{L}(r(u, i \mid \Theta), r(u, j \mid \Theta))],
\end{equation}
where $P_{ns}(j \mid \Theta )$ denotes the negative sampling probability that a negative item is drawn. Departing from conventional paradigms, we introduce the PNN learning framework, which stratifies unlabeled data into neutral and negative classes, employing set-level rankings for parameter optimization:
\begin{equation}
\label{eq:pnn}
\max_\Theta \sum_{u \in U} \Sigma_{pos}^u >_u \Sigma_{neu}^u >_u \Sigma_{neg}^u.
\end{equation}
As a novel paradigm, our objective is to theoretically dissect the impact of PNN on recommendation performance. Recent research reveals the correlation between OPAUC and top-K evaluation metrics~\cite{SCF23}, motivating our investigation into PNN efficacy via this metric. OPAUC is expressed as:
\begin{equation}
\begin{gathered}
\widehat{OPAUC}(\Theta, \gamma, \delta)=\frac{1}{|U|} \sum_{u \in U} \int_\gamma^\delta TPR_{u, \theta}\left[FPR_{u, \theta}^{-1}(s)\right] \mathrm{d}s,
\end{gathered}
\end{equation}
where $TPR$ and $FPR$ represent true positive rates and false positive rates, respectively. OPAUC only cares about the performance within a given false positive rate (FPR) range $[\gamma, \delta]$. In line with prior works~\cite{SCF23}, we consider the special case of OPAUC with $\gamma=0$, yielding the non-parametric estimator:
\begin{equation}
\begin{gathered}
\widehat{OPAUC}(\delta)=\frac{1}{|U|} \sum_{u \in U} \frac{1}{|\mathcal{I}_u^{+}|} \frac{1}{|\mathcal{I}_u^{-}|} \sum_{i \in I_u^{+}} \sum_{j \in S_{I_u^{-}}\left[1, |\mathcal{I}_u^{-}|\cdot \delta\right]} \mathbb{I}\left(r_{ui}>r_{uj}\right),
\end{gathered}
\end{equation}
where $I(\cdot)$ is an indicator function and $S_{I_u^{-}}\left[1, n_{-} \cdot \delta\right]$ denotes the subset of the top-ranked negative items(subsequently abbreviated as $S_{I_u^{-}}$). However, as previously mentioned, given that unlabeled data harbors signals beyond mere negativity, a chasm emerges between $\mathcal{I}_u^{-}$ and $\mathcal{I}_u^{un}$. Our optimization endeavor cannot directly procure top-ranked negative items. Thus, we invoke further refinement:
\begin{equation}
\begin{gathered}
\widehat{OPAUC}(\delta)=|K|\sum_{i \in I_u^{+}} \sum_{i^n \in S_{I_u^{un}}} \mathbb{I}\left(r_{ui}>r_{uj}\right)\mathbb{I}(i^n \in S_{I_u^{-}} ).
\end{gathered}
\end{equation}
$|K|$ denotes the constant unrelated to derivation. And for brevity, we've omitted the user. Since direct sampling from $\mathcal{I}_u^{-}$ is unfeasible, we initially sample from unlabeled data and subsequently utilize an indicator function to identify top-ranked negative samples. Then, for these samples, it is evident that $r_{ui^n} >r_{uj}$. Consequently, we can rewrite the equation as
\begin{equation}
\begin{gathered}
\label{eq:evidence}
\widehat{OPAUC}(\delta)= |K|\sum_{i \in I_u^{+}} \sum_{i^n \in I_u^{un}} \sum_{j \in I_u^{un}} \mathbb{I}\left(r_{ui}>r_{ui^n} >r_{uj}\right)\mathbb{I}(i^n \in S_{I_u^{-}})
\end{gathered}
\end{equation}
Obviously, by expressing Eq.~\ref{eq:evidence} in set notation and introducing symbols for positive, neutral, and negative classes within PNN, we arrive at Eq.~\ref{eq:pnn}.

It becomes evident that neutral samples and the set-level rankings introduced by PNN can directly optimize the OPAUC metric, thereby enhancing various evaluation metrics of recommender systems. Simultaneously, through Eq.~\ref{eq:evidence}, it's apparent that the concrete implement PNN needs to grapple with two key challenges. First, $\mathbb{I}(i^n \in S_{I_u^{-}})$ implies the necessity to devise a reliable scheme to distinguish neutral samples and negative ones. Second, $\mathbb{I}\left(r_{ui}>r_{ui^n} >r_{uj}\right)$ necessitates an approach beyond conventional positive-negative learning paradigms.


%% file: 4.methodology.tex
\section{Methodology}
\begin{figure*}[h]
\setlength{\abovecaptionskip}{3pt plus 2pt minus 2pt}
  \centering
  \includegraphics[width=0.8\textwidth]{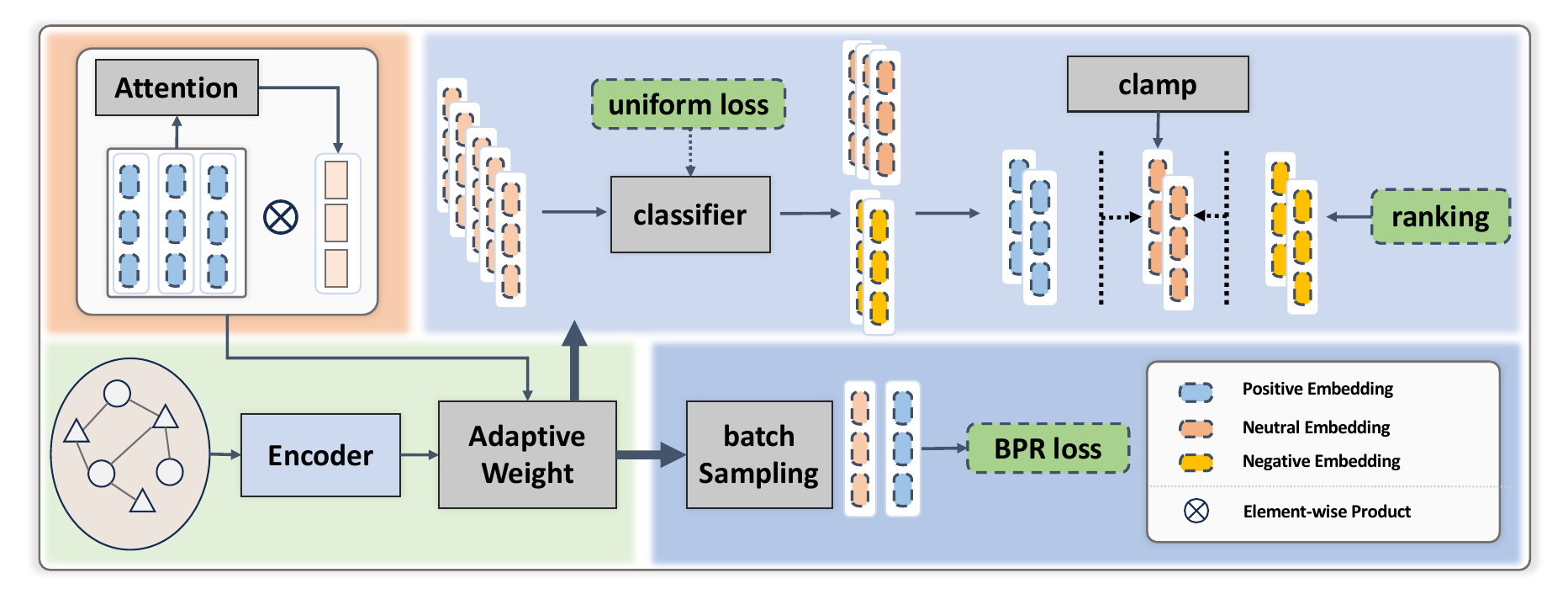}
  \caption{The overall workflow of our PNN learning paradigm.}
  \label{fig:model}
  \vspace{-3mm}
\end{figure*}
Driven by the aforementioned limitations, we propose a novel generic positive-neutral-negative (PNN) learning paradigm. In the following sections, we introduce a concrete implementation as a loss function $\mathcal{L}$, which allows it to be seamlessly integrated with many mainstream CF models. The implementation boils down to two key components: a classifier that reliably distinguishes between neutral and negative items and a ranking mechanism that can effectively handle set-level triple-wise ranking relationships. The workflow is illustrated in Figure~\ref{fig:model}. Note that these steps can be implemented by different methods and thus the overall paradigm is generic.

\subsection{Semi-Supervised Learning}
The key challenge of the classification is a lack of supervised signals that are essential to separate neutral and negative items. Inspired by pseudo-labelling techniques~\cite{WWS22, RDR21}, we design a novel semi-supervised learning method that adequately leverages the knowledge acquired from the BPR loss as the decision foundation. Given user $u \in U$, we formulate the loss function as follows:
\begin{equation}
\label{L}
\mathcal{L}_u = (1 - \lambda) \mathcal{L}_{\mathrm{BPR}} + \lambda\mathcal{L}_{\mathrm{PNN}},
\end{equation}
where $\mathcal{L}_{\mathrm{BPR}}$ represents the BPR loss, $\mathcal{L}_\mathrm{PNN}$ denotes the PNN loss that factors in the influence of both the classification and ranking tasks, and $\lambda$ is a key parameter that allows us to adaptively adjust the weights between the BPR loss and the PNN loss.

At the beginning of the training process, we mainly rely on the BPR loss to train the model to acquire prior knowledge about user preferences via
\begin{equation}
\begin{split}
\begin{aligned}
\mathcal{L}_{\mathrm{BPR}}&= -\ln [\sigma(s(\mathbf{e}_u ,\mathbf{e}_{i}) - s(\mathbf{e}_u ,\mathbf{e}_{j}))],
\end{aligned}
\end{split}
\end{equation}
where $s(\cdot,\cdot)$ denotes the score function (e.g., inner product) used to calculate similarities, $\sigma(\cdot)$ denotes the sigmoid function, and $\mathbf{e}_u$, $\mathbf{e}_i$, and $\mathbf{e}_j$ are user/item embeddings. Here $i$ is a positive sample with which the user interacted, and $j$ is a negative sample selected from unlabeled data. 

We use dynamic negative sampling (DNS)~\cite{ZCW13}, a hard negative sampling method, to select an item $j$ with probability proportional to $p_{j} \propto s(\mathbf{e}_u ,\mathbf{e}_{j})$. The intuition of DNS is to select negative items that are ``closer'' to the user. Such negative samples can yield greater gradients~\cite{RF14,CLJ22}, thereby accelerating the training process. This choice is important because we expect the model to quickly acquire valuable knowledge from the BPR loss and then start to optimize $\mathcal{L}_{\mathrm{PNN}}$. However, this strategy needs to calculate the scores of all unlabeled data to assign probabilities. The time complexity is prohibitive. To balance efficiency and effectiveness, we propose to select the negative sample by batch style via
\begin{equation}
\mathbf{e}_{j}=\underset{n \in {batch} \backslash i}{\arg \max }s(\mathbf{e}_u ,\mathbf{e}_{n}).
\end{equation}

Diverging from the approach of selecting negative samples from the entire item set, we restrict the scope to a single batch. Within this batch, we directly identify the highest-score sample as the negative sample, excluding the current positive sample. 
The parameter $\lambda$ is designed to indicate the readiness to make use of the PNN loss. A conventional method is to set $\lambda$ as a hyperparameter or learn a function of the number of completed training epochs~\cite{WWS22}. However, such methods are not aligned with our objective. More specifically, once the current model has acquired sufficient knowledge for the classification task, we want to reduce the weight of the BPR loss and start to focus on optimizing the PNN loss. Conversely, when the classification task is arduous, we may want to increase the weight of the BPR loss.

Similarly, how to design $\lambda$ is also a challenging task due to a lack of supervised signals. Inspired by previous works~\cite{DQY20,MZW21}, we propose a user-aware attention model to indirectly assess the classification performance based on the user's 
positive items sets $\Sigma_{pos}^u$.
Intuitively, if the model can correctly identify these items as positive samples, it suggests that the model has adequately learned user preferences from the BPR loss. Following this intuition, we devise $\lambda$ as follows:
\begin{equation}
\lambda =\sigma[\textbf{W}_1(\sum_{i \in \Sigma_{pos}^u} \alpha_{i}^{attr} \mathbf{e}_{i})],
\end{equation}
where $\textbf{W}_1 \in \mathbb{R}^{1\times d}_{+}$ is a trainable matrix, and the sigmoid function $\sigma(\cdot)$ is used to map values to $(0, 1)$. Here $\alpha_{i}^{attr}$ represents the aggregation weight, which can be computed via
\begin{equation}
\begin{aligned}
\alpha_{i}^{attr} &= \frac{\exp \left(\beta_i^{attr}\right)}{\sum_{j \in \Sigma_{pos}^u} \exp \left(\beta_j^{attr}\right)},\\
\beta_{i}^{attr}&= \textbf{e}_{u}^{T} \tanh (\textbf{W}_2 \textbf{e}_{i}+\textbf{b}),
\end{aligned}
\end{equation}
where $\textbf{W}_2 \in \mathbb{R}^{d\times d}_{+}$ and $\textbf{b} \in \mathbb{R}^{d\times 1} $ are trainable parameters, and the user embedding $\textbf{e}_u$ is used as a query vector. In this way, the attention mechanism can serve as an indirect evaluation metric for classification performance. 

When our model possesses sufficient knowledge, the positive items should exhibit a high degree of similarity to the user's preferences. As a result, $\alpha_{i}^{attr}$ will receive a higher value. Consequently, the weight $\lambda$ will also increase, prioritizing the significance of $\mathcal{L}_{\mathrm{PNN}}$ in the overall loss function.

\subsection{PNN Loss Function}
After acquiring sufficient knowledge from the BPR loss to warm-start the PNN loss, we can now focus on classifying unlabeled data into either neutral or negative and effectively handling set-level triple-wise ranking relationships. 

\subsubsection{Classification Task}
Following the convention of implicit CF, for a given user $u \in U$, the positive class $\Sigma_{pos}^u$ consists of all her interacted items. Let $|\Sigma_{pos}^u| = N_u$. The classification between neutral and negative is achieved by the PNN loss function. 

Intuitively, the items with the lowest similarities to $u$ are more likely to be truly negative. Driven by this intuition, we take a greedy approach to iteratively select $N_u$ negative items. Since the knowledge gained from the semi-supervised learning process could still be limited, our design principle is to put only truly reliable negative items into $\Sigma_{neg}^u$. For this reason, we limit the size of $\Sigma_{neg}^u$ to $N_u$ to reduce the risk of introducing false negative samples. Note that $N_u$ is much smaller than the number of unlabeled data. Formally, in each iteration, we update $\Sigma_{neg}^u$ as follows:
\begin{equation}
\begin{aligned}
i^{neg} &= \underset{i \in \mathcal{E}}{\arg \min}\ s(\mathbf{e}_u,\mathbf{e}_i),\\
\mathcal{E}^u &= \mathcal{E}^u - i^{neg},\\
\Sigma_{neg}^u &= \Sigma_{neg}^u \cup \{i^{neg}\},
\end{aligned}
\end{equation}
where $\mathcal{E}$ is initialized to the set of $u$'s unlabeled items $\mathcal{E}^u$, and $\Sigma_{neg}^u$ is initialized to $\varnothing$. Consequently, the neutral class consists of all remaining unlabeled data, namely $\mathcal{E}^u - \Sigma_{neg}^u$.

Recalling that the PNN loss function is optimized on top of the knowledge acquired from the BPR loss. Since the BPR loss is designed to set positive items apart from unlabeled data , it may lead to clusters of unlabeled data with similar scores, which is detrimental to our classification goal. Considering the scenario where both $i_1$ and $i_2$ receive identical scores of 2 points, in such a case, discerning which is more likely to represent a negative sample becomes challenging. To address this problem, it is natural to impose a constraint that unlabeled data are sufficiently distant from each other. Specifically, we introduce the following loss factor: 
\begin{equation}
\mathcal{L}_{\mathrm{uniform}} =  \log \underset{i, i^{\prime} \in \mathcal{E}^u}{\mathbb{E}} e^{-2\|\mathbf{e}_{{i}} -\mathbf{e}_{{i^{\prime}}}\|^2 }.
\label{eq:uni}
\end{equation}
By utilizing the Euclidean distance, we encourage unlabeled data to push each other further apart, leading to a more uniform distribution~\cite{WI20}. As a result, we can more reliably distinguish between neutral and negative items.

\subsubsection{Ranking Task}
Next, we discuss how to handle set-level triple-wise ranking relationships so that our solution can be seamlessly integrated with mainstream CF models. It is important to realize that preference ranking may not satisfy the transitive property~\cite{TA69,EI12,MG17}. Hence we need to decompose the set-level triple-wise ranking relationship $\Sigma_{pos}^u >_u \Sigma_{neu}^u >_u \Sigma_{neg}^u$ into three pairwise ranking relationships to avoid information loss: $\Sigma_{pos}^u >_u \Sigma_{neg}^u$, $\Sigma_{pos}^u >_u \Sigma_{neu}^u$, and $\Sigma_{neu}^u >_u \Sigma_{neg}^u$. While the BPR loss operates at the item level, it can still well enforce $\Sigma_{pos}^u >_u \Sigma_{neg}^u$. So the PNN loss should take into consideration the latter two pairwise rankings.

Once we have decomposed pairwise rankings, we move to address the challenge due to the set-level formulation. To this end, we propose a two-step centroid ranking approach to convert set-level relationships into item-level relationships while maintaining the desirable property of set-level rankings. In the first step, we propose to represent the set of items in each of the three classes by its centroid, namely $\mathbf{e}_{i^{neu}} = \underset{i \in \Sigma_{neu}^u}{\mathbb{E}}\!\!\!\mathbf{e}_{i}$, $\mathbf{e}_{i^{+}} = \underset{i \in \Sigma_{pos}^u}{\mathbb{E}}\!\!\!\mathbf{e}_{i}$, and $\mathbf{e}_{i^{-}} = \underset{i \in \Sigma_{neg}^u}{\mathbb{E}}\!\!\!\mathbf{e}_{i}$. This natural idea also gains another benefit of mitigating the impact of class imbalance: the size of $\Sigma_{neu}^u$ is much larger than that of $\Sigma_{pos}^u$ or $\Sigma_{neg}^u$. Then we can capture the two pairwise rankings via
\begin{equation}
\begin{aligned}
\mathcal{L}_{\mathrm{rank}} = &-\ln [\sigma(s(\mathbf{e}_u ,\mathbf{e}_{i^+}) - s(\mathbf{e}_u ,\mathbf{e}_{i^{neu}}))] \\ &-\ln [\sigma(s(\mathbf{e}_u ,\mathbf{e}_{i^{neu}}) - s(\mathbf{e}_u ,\mathbf{e}_{i^-}))].
\end{aligned}
\label{eq:pg}
\end{equation}
Here we construct $\mathcal{L}_{\mathrm{rank}}$ in a format similar to the BPR loss. This is because it guarantees that the magnitudes of $\mathcal{L}_{\mathrm{BPR}}$ and $\mathcal{L}_{\mathrm{rank}}$ remain comparable and that the learning process would not be dominated by one of them~\cite{SZC20,YPW23}. 

Unfortunately, the above method cannot guarantee that every item in a class ranks higher than any item in another class. Take positive and neutral as examples, an ideal ranking need to satisfy $\forall i \in \Sigma_{pos}^u >_u \forall j \in \Sigma_{neu}^u $.
It is not hard to understand because Eq.~\ref{eq:pg} can only enforce the ranking relationships on the centroids. Some of the classes may overlap with each other. Therefore, we want to maintain larger margins between different classes so that the set-level property can be respected.
\begin{figure}[h]
  \centering
\includegraphics[width=0.7\linewidth]{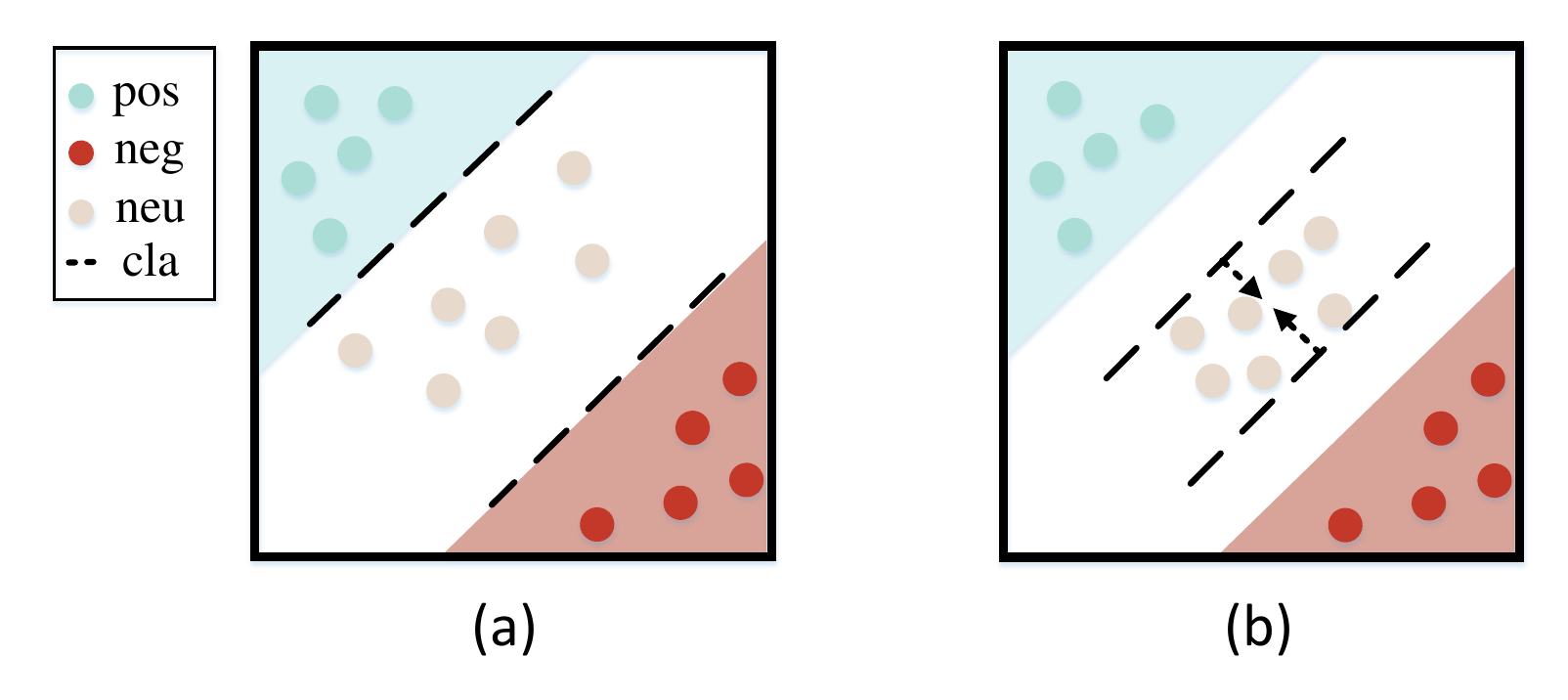}
\vspace{-3mm}
  \caption{The overall process of the clamp mechanism. pos, neu, and neg, respectively represent the positive, neutral, and negative class, and cla denotes clamp embedding.}
  \label{fig:clamp}
  \vspace{-3mm}
\end{figure}
Intuitively, we can set a fixed margin between classes as usual to avoid overlap. However, we have no prior knowledge to determine the positive-neutral-negative reasonable margin, artificially setting the margin will introduce inductive bias. Inspired by previous works~\cite{GSS14, HHD18, YYX22, YXC23}, in the second step, we propose a novel clamp mechanism to adaptively generate the margin. Our insight is to utilize two clamp embeddings to tightly constrain the neutral classes in the embedding space. Since the neutral class occupies a position between the positive and negative classes, clamping the neutral class creates a distinct margin with the other two classes. As a result, this approach enables better separation and avoids overlap among the different classes. Figure~\ref{fig:clamp} vividly shows this process.

Our clamp embedding requires the presence of one embedding between positive and neutral, as well as another embedding between neutral and negative. For the sake of brevity, we will focus solely on the positive-related embedding. First, we generate noise $\Delta_{pos} \in \mathbb{R}^{d}$ as follows. We adopt a uniform distribution on the interval $[0, 0.1]$ to decide the magnitude. A uniform distribution introduces a certain amount of randomness, which is also beneficial to improve the model's robustness. We fix the direction of $\Delta_{pos}$ towards the positive. Take the direction towards the positive class as an example. We can generate the clamp embedding as
\begin{equation}
\begin{gathered}
\Delta_{pos}=\bar{\Delta}_{pos} \odot \frac{\left(\mathbf{e}_{i^{+}}\right)}{||\left(\mathbf{e}_{i^{+}}\right)||}, \bar{\Delta}_{pos} \in \mathbb{R}^d \sim U(0,0.1), \\ 
\mathbf{e}_{clamp}^{pos} = \mathbf{e}_{i^{neu}} + \Delta,
\end{gathered}
\end{equation}
where $\odot$ is the element-wise product, and $f(x) = \frac{x}{||x||}$ is the sign function to determine the direction vector of the noise. $\mathbf{e}_{clamp}^{neg}$ can be obtained in a similar way. 

Note that we want to enlarge the margin by tightly constraining the neutral classes in the embedding space. For this reason, we minimize the distance between the two clamp embeddings, compressing the neutral class similar to how a clamp operates. which leads to the following:
\begin{equation}
\label{eq:clamp}
\mathcal{L}_{\mathrm{constrain}} = {\mathbb{E}} \| (\mathbf{e}_{clamp}^{pos} - \mathbf{e}_{clamp}^{neg} ) \|^2.
\end{equation}
It is worth noting that Eq.~\ref{eq:clamp} can be simplified, providing an alternative interpretation of the clamp mechanism. However, to ensure readers' understanding, we retain the original form.
Now we are ready to present the PNN loss $\mathcal{L}_{\mathrm{PNN}}$, which accounts for both classifying unlabeled data and tackling set-level triple-wise rankings:
\begin{equation}
\begin{aligned}
\mathcal{L}_{\mathrm{PNN}} = \alpha\mathcal{L}_{\mathrm{constrain}} +\beta\mathcal{L}_{\mathrm{uniform}} +\mathcal{L}_{\mathrm{rank}},
\end{aligned}
\end{equation}
where $\alpha$ and $\beta$ are hyperparameters to control the relative weights of $\mathcal{L}_{\mathrm{constrain}}$ and $\mathcal{L}_{\mathrm{uniform}}$, respectively.
\subsection{Mini-Batch Training}
In this section, we explain how PNN can be easily integrated with different mainstream CF models. Ideally, we want PNN to use all observed and unobserved items of a given user in each training epoch. In this case, the final loss function $\mathcal{L}$ can be expressed as $\mathcal{L} = \sum_{u \in U} \mathcal{L}_u$. However, for efficiency reasons, mainstream CF models generally adopt mini-batch training, where a mini-batch consists of user-item interactions in the form of $(u,i)$. Thus, accessing all user interaction data in a mini-batch is impossible. 

To seamlessly integrate PNN, we propose to relax the ranking as $i>_u\Sigma_{neu}^{\widetilde{u}}>_u \Sigma_{neg}^{\widetilde{u}}$, where $\Sigma_{neu}^{\widetilde{u}}$ and $\Sigma_{neg}^{\widetilde{u}}$ denote the neutral class and negative class that can be built in mini-batch training, respectively. Note that $|\Sigma_{neu}^{\widetilde{u}}| \ll |\Sigma_{neu}^{u}|$ and $|\Sigma_{neg}^{\widetilde{u}}| \ll |\Sigma_{neg}^{u}|$. If a model can provide unlabeled data, we can directly utilize them to construct $\Sigma_{neu}^{\widetilde{u}}$ and $\Sigma_{neg}^{\widetilde{u}}$. However, when such data is not available, we design a scheme with reference to contrastive learning~\cite{WWF21,YYX22}. We utilize items that were interacted with by other users, but not user $u$, in a mini-batch as unobserved items to form neutral and negative classes. Thus we have $\mathcal{L} = \sum_{\substack{(u, i) \in D \\ }} \mathcal{L}_u$. Similar to contrastive learning, during training, when $i$ takes all the values in $\Sigma_{pos}^u$, we can get the following approximation:
\begin{equation}
\begin{aligned}
\prod_ {i \in \Sigma_{pos}^u}p( i >_u\Sigma_{neu}^{\widetilde{u}}>_u\Sigma_{neg}^{\widetilde{u}}) \approx p(\Sigma_{pos}^u>_u\Sigma_{neu}^u>_u\Sigma_{neg}^u).
\end{aligned}
\end{equation}
This approximation enables PNN to be seamlessly applied to almost all mainstream CF models. Specifically, we just need to replace the original pairwise ranking loss function with $\mathcal{L} = \sum_{\substack{(u, i) \in D \\}} \mathcal{L}_u$.

%% file: 5.experiment.tex
\section{EXPERIMENTS}
\label{experiment}

\begin{table}[t]
\caption{The statistics of the datasets used in the experiments.}
\begin{center}
\begin{tabular}{@{}lccccc@{}}
\toprule
\textbf{Dataset} & \textbf{Users} & \textbf{Items} & \textbf{Interactions} & \textbf{Sparsity}\\ \midrule
ML-1M & 6,041 & 3,707 & 1,000,209 & 95.53\%\\ 
Yelp & 31,669 & 38,049 & 1,561,406 & 99.87\% \\ 
Gowalla & 29,859 & 40,982 & 1,027,370 & 99.92\% \\
Foursquare & 1,084 &38,334 & 91,024 & 99.78\%\\
\bottomrule
\end{tabular}
\label{tab:statistics}
\end{center}
\vspace{-3mm}
\end{table}

\begin{table*}[t]
\caption{Experimental results of different CF models with (w) or without (w/o) the PNN loss function. The best results are boldfaced, and the improvement is the average of all metrics.}
\centering
\begin{tabular}{@{}clllllllllllll@{}}
\toprule
\multirow{2}{*}{\textbf{Datasets}} & \multirow{2}{*}{\textbf{Metric}} & \multicolumn{2}{c}{\textbf{BPR-MF}} & \multicolumn{2}{c}{\textbf{LightGCN}} & \multicolumn{2}{c}{\textbf{NGCF}} & \multicolumn{2}{c}{\textbf{SGL}} &  \\ \cmidrule(l){3-10} 
 &  & w/o & w & w/o & w & w/o & w & w/o & w  \\ \midrule
\multirow{6}{*}{ML-1M} 
 & Recall@10 & 0.1635 & \textbf{0.1840} & 0.1730 & \textbf{0.1971} & 0.1614 & \textbf{0.1880} & 0.1769 & \textbf{0.1942}  \\
 & Recall@20 & 0.2502 & \textbf{0.2742} & 0.2647 & \textbf{0.2909} & 0.2475 & \textbf{0.2839} & 0.2665 & \textbf{0.2887}  \\
 & Hit@10 & 0.7419 & \textbf{0.7753} & 0.7603 & \textbf{0.7997} & 0.7407 & \textbf{0.7803} & 0.7699 & \textbf{0.7929}  \\
 & Hit@20 & 0.8374 & \textbf{0.8613} &  0.8546 & \textbf{0.8811} & 0.8359 & \textbf{0.8694} & 0.8591 & \textbf{0.8752}  \\
 & NDCG@10 & 0.2554 & \textbf{0.2835} & 0.2682 & \textbf{0.2895} & 0.2541 & \textbf{0.2898} & 0.2682 & \textbf{0.2909} \\
 & NDCG@20 & 0.2619 & \textbf{0.2869} & 0.2749 & \textbf{0.2951} & 0.2595 & \textbf{0.2951} & 0.2755 & \textbf{0.2959}  \\
\cmidrule(l){2-10} 
  & Improvement &  & \textbf{8.34\%} &  & \textbf{7.90\%} &  & \textbf{11.39\%} &  & \textbf{6.47\%} &    \\ \midrule 

\multirow{6}{*}{Gowalla} 
 & Recall@10 & 0.0938 & \textbf{0.1236} & 0.1280 & \textbf{0.1357} & 0.1011 & \textbf{0.1215} & 0.1331 & \textbf{0.1437}  \\
 & Recall@20 & 0.1389 & \textbf{0.1787} & 0.1849 & \textbf{0.1972} & 0.1490 & \textbf{0.1813} & 0.1897 & \textbf{0.2085}  \\
 & Hit@10 & 0.1989 & \textbf{0.2534} & 0.2553 & \textbf{0.2712} & 0.2110 & \textbf{0.2491} & 0.2667 & \textbf{0.2846}  \\
 & Hit@20 & 0.2763 & \textbf{0.3366} & 0.3432 & \textbf{0.3635} & 0.2900 & \textbf{0.3414} & 0.3519 & \textbf{0.3797}  \\
 & NDCG@10 & 0.0674 & \textbf{0.0896} & 0.0917 & \textbf{0.0981} & 0.0722 & \textbf{0.0875} & 0.0966 & \textbf{0.1044} \\
 & NDCG@20 & 0.0804 & \textbf{0.1052} & 0.1079 & \textbf{0.1157} & 0.0859 & \textbf{0.1046} & 0.1128 & \textbf{0.1229}  \\
\cmidrule(l){2-10} 
  & Improvement &  & \textbf{28.90\%} &  & \textbf{6.50\%} &  & \textbf{20.09\%} &  & \textbf{8.25\%} &    \\ \midrule 

\multirow{6}{*}{Yelp} 
 & Recall@10 & 0.0427 & \textbf{0.0547} & 0.0543 & \textbf{0.0668} & 0.0439 & \textbf{0.0587} & 0.0612 & \textbf{0.0674}  \\
 & Recall@20 & 0.0717 & \textbf{0.0873} & 0.0884 & \textbf{0.1079} & 0.0738 & \textbf{0.0946} & 0.0990 & \textbf{0.1084}  \\
 & Hit@10 & 0.1537 & \textbf{0.1917} & 0.1896 & \textbf{0.2262} & 0.1593 & \textbf{0.2047} & 0.2101 & \textbf{0.2275}  \\
 & Hit@20 & 0.2398 & \textbf{0.2819} & 0.2847 & \textbf{0.3317} & 0.2460 & \textbf{0.3014} & 0.3098 & \textbf{0.3334}  \\
 & NDCG@10 & 0.0334 & \textbf{0.0435} & 0.0432 & \textbf{0.0539} & 0.0345 & \textbf{0.0469} & 0.0487 & \textbf{0.0540} \\
 & NDCG@20 & 0.0432 & \textbf{0.0543} & 0.0546 & \textbf{0.0676} & 0.0447 & \textbf{0.0588} & 0.0614 & \textbf{0.0677}  \\
\cmidrule(l){2-10} 
  & Improvement &  & \textbf{24.68\%} &  & \textbf{21.58\%} &  & \textbf{30.07\%} &  & \textbf{9.44\%} &    \\ \midrule 

\multirow{6}{*}{Foursquare} 
 & Recall@10 & 0.0262 & \textbf{0.0332} & 0.0369 & \textbf{0.0376} & 0.0325 & \textbf{0.0356} & 0.0358 & \textbf{0.0366}  \\
 & Recall@20 & 0.0381 & \textbf{0.0472} & 0.0537 & \textbf{0.0554} & 0.0474 & \textbf{0.0542} & 0.0524 & \textbf{0.0525}  \\
 & Hit@10 & 0.1717 & \textbf{0.2041} & 0.2364 & \textbf{0.2364} & 0.2151 & \textbf{0.2225} & 0.2281 & \textbf{0.2299}  \\
 & Hit@20 & 0.2336 & \textbf{0.2770} & 0.3158 & \textbf{0.3250} & 0.2973 & \textbf{0.3186} & \textbf{0.3213} & {0.3130}  \\
 & NDCG@10 & 0.0283 & \textbf{0.0352} & 0.0381 & \textbf{0.0398} & 0.0345 & \textbf{0.0392} & 0.0391 & \textbf{0.0392} \\
 & NDCG@20 & 0.0328 & \textbf{0.0405} & 0.0443 & \textbf{0.0466} & 0.0403 &  \textbf{0.0469} & \textbf{0.0455} & {0.0449}  \\
\cmidrule(l){2-10} 
  & Improvement &  & \textbf{22.65\%} &  & \textbf{2.93\%} &  & \textbf{10.74\%} &  & \textbf{-} &    \\ \midrule 
\end{tabular}
\label{tab:Overall-comprehension-1}
\vspace{-3mm}
\end{table*}
In this section, we conduct comprehensive experiments to answer the following key research questions:
\begin{itemize}
\item \textbf{RQ1:} How does integrating PNN help different mainstream CF models boost their performance? How does the performance of PNN  compare to state-of-the-art baselines?
\item \textbf{RQ2:} How does the neutral class affect model performance?
\item \textbf{RQ3:} How do different components of PNN affect model performance? 
\item \textbf{RQ4:} How do different parameter settings affect model performance? 
\item \textbf{RQ5:} Is PNN efficient for practical use?
\end{itemize}

\begin{table*}[]
 \caption{The performance comparison between PNN and other state-of-the-art methods.}
 \vspace{-2mm}
\label{tab:Overall-comprehension-2}
\begin{tabular}{@{}llccccccc@{}}
\toprule
\textbf{Dataset} & \textbf{Methods} & HR@10 & HR@20 & Recall@10 & Recall@20 & NDCG@10 & NDCG@20 & MRR@10 \\ \midrule
\multirow{5}{*}{ML-1M} 
  & \textbf{SRNS} & 0.7611 & 0.8533 & 0.1746 & 0.2618 & 0.2639 &  0.2687 & 0.4661 \\
 & \textbf{MixGCF} & 0.7601 & 0.8483 & 0.1739 & 0.2604 & 0.2657 & 0.2702 & 0.4680 \\
 & \textbf{UIB} & 0.7346 & 0.8318 & 0.1559 & 0.2406 & 0.2453 & 0.2520 & 0.4279\\ 
& \textbf{SimpleX} & 0.7586 & 0.8518 & 0.1709 & 0.2573 & 0.2626 & 0.2675 & 0.4607 \\ 
& \textbf{ANS} & 0.7675 & 0.8583 & 0.1759 & 0.2625 & 0.2624 & 0.2683 & 0.4640 \\
& \textbf{PNN} & \textbf{0.7753}& \textbf{0.8613} & \textbf{0.1840} & \textbf{0.2742} & \textbf{0.2835 }& \textbf{0.2869} & 
\textbf{0.4837 }\\ 
 
 \midrule
 
\multirow{5}{*}{Gowalla} 
  & \textbf{SRNS} & 0.2295 & 0.3177 & 0.1171 & 0.1711 & 0.0814 & 0.0972 & 0.1042 \\
 & \textbf{MixGCF} & 0.2401 & 0.3254 & 0.1189 & 0.1728 & 0.0853 & 0.1008 & 0.1145 \\
 & \textbf{UIB} & 0.2275 & 0.3161 & 0.1140 & 0.1695 & 0.0794 & 0.0954 & 0.1029\\ 
& \textbf{SimpleX} & 0.2192 & 0.3082 & 0.1074 & 0.1617 & 0.0736 & 0.0892 & 0.0949 \\
& \textbf{ANS} & 0.2467 & 0.3317 & 0.1206 & 0.1756 & 0.0869 & 0.1027 & 0.1180 \\ 
& \textbf{PNN} & \textbf{0.2534} & \textbf{0.3366} & \textbf{0.1236} & \textbf{0.1787} & \textbf{0.0896} &\textbf{0.1052} & 
\textbf{0.1217} \\ 
 \midrule

 \multirow{5}{*}{Foursquare} 
  & \textbf{SRNS} & 0.1948 & 0.2596 & 0.0314 & 0.0418 & 0.0351 & 0.0392& 0.0840 \\
 & \textbf{MixGCF} & 0.1994 & 0.2752 & 0.0316 & 0.0458 & 0.0351 & 0.0405 & 0.0863 \\

 & \textbf{UIB} & 0.1662 & 0.2299 & 0.0265 & 0.0375 & 0.0280 & 0.0322 & 0.0717\\ 
& \textbf{SimpleX} & 0.1801 & 0.2650 & 0.0267 & 0.0421 & 0.0267 & 0.0329 & 0.0641 \\ 
& \textbf{ANS} & 0.1958 & 0.2798 & 0.0303 & 0.0461 & 0.0321 & 0.0386 & 0.0821 \\ 
& \textbf{PNN} & \textbf{0.2041} & \textbf{0.2770} & \textbf{0.0332} & \textbf{0.0472} & \textbf{0.0352} & \textbf{0.0405} & 
\textbf{0.0876}\\ 
 
 \midrule
 
\multirow{5}{*}{Yelp} 
  & \textbf{SRNS} & 0.1834 & 0.2779 & 0.0520 & 0.0862 & 0.0406 & 0.0522 & 0.0690 \\
 & \textbf{MixGCF} & 0.1891 & 0.2781 & 0.0538 & 0.0871 & 0.0424 & 0.0536 & 0.4680 \\
 & \textbf{UIB} & 0.1592 & 0.2504 & 0.0445 & 0.0756 & 0.0350  & 0.0455 & 0.0598\\ 
& \textbf{SimpleX} & 0.1850 & 0.2780 & 0.0519 & 0.0861 & 0.0410 & 0.0524 & 
0.0706 \\ 
& \textbf{ANS} & 0.1844 & 0.2777 & 0.0524 & 0.0862 & 0.0414 & 0.0527 & 
0.0708 \\ 
& \textbf{PNN} & \textbf{0.1917} & \textbf{0.2819} & \textbf{0.0547} & \textbf{0.0873} & \textbf{0.0435} & \textbf{0.0543} & \textbf{0.0739} \\ 

 \bottomrule
 \end{tabular}
\label{tab:Overall-comprehension-2}
 \vspace{-3mm}
\end{table*}

\subsection{Experimental Setup}
\subsubsection{Datasets}
We conduct experiments on four datasets widely used in the literature to evaluate the performance of PNN: (1) \textbf{MovieLens}: It contains user ratings on movies. We use 1M versions and treat all rating movies as interactive items. (2) \textbf{Yelp}: It contains user reviews of restaurants and bars. We use the transaction records after Jan. 1st, 2018. (3) \textbf{Gowalla}: This is a check-in dataset obtained from Gowalla, where users share their locations by checking-in. (4) \textbf{Foursquare}: This dataset contains check-ins in NYC and Tokyo collected for about 10 months. We split the historical interactions into training, validation, and test sets in a ratio of 8:1:1. Table~\ref{tab:statistics} summarizes the statistics of the datasets. 
\subsubsection{Implementation Details}
The experiments are conducted on the NVIDIA Ampere A100-40G GPU using PyTorch. To ensure reproducibility, we implement all methods by using the RecBole v1.1.1 framework~\cite{ZMH21}. We maintain a fixed size of 64 for the embeddings. The optimization of parameters is carried out using Adam with a default learning rate of 0.001 and a default mini-batch size of 2048. The $L_2$ regularization coefficient is set to $10^{-4}$. We evaluate the performance of PNN by varying the value $\alpha$ within the range of $[0,1]$ and the value $\beta$ within the range of $[0,1]$. We meticulously tune all hyper-parameters on the validation datasets and report the best performance for all basic models. 
\subsubsection{Evaluation Protocols}
We evaluate performance employing standard metrics for top-K recommendations, encompassing Normalized Discounted Cumulative Gain (NDCG@K), Hit Rate (HR@K), Recall (Recall@K), and Mean Reciprocal Rank (MRR). Our findings are based on the average results obtained from five independent runs and we conduct statistical significance analysis by calculating $p$-values against the best-performing baseline.

\subsubsection{Baseline}
To validate the effectiveness of our solution, we integrate PNN with a wide range of mainstream CF models.
\begin{itemize}
    \item \textbf{MF-BPR}~\cite{RFG12}, which is the most classic matrix factorization.
    \item \textbf{NGCF}~\cite{WHW19}, which employs a graph neural network to leverage high-order information.
    \item \textbf{LightGCN}~\cite{HDW20}, which eliminates the redundant components from NGCF to improve performance.
    \item \textbf{SGL}~\cite{WWF21}, which incorporates contrastive learning to improve the accuracy and robustness.
\end{itemize}
Furthermore, we compare PNN with several representative state-of-the-art methods. Following the conventional setting~\cite{RFG12,WYZ17,CLJ22}, we consider BPR-MF as the basic model of the following methods.
\begin{itemize}
    \item \textbf{SimpleX}~\cite{MZW21}, which extracts a large volume of high-information samples from unlabeled data to serve as negative samples and uses cosine similarity for ranking.
    \item \textbf{UIB}~\cite{ZZY22}, which introduces a personalized boundary to penalize samples crossing the threshold.
    \item \textbf{SRNS}~\cite{DQY20}, which selects hard samples from unlabeled data to serve as negative samples while incorporating a variance mechanism to filter out false negative samples.
    \item \textbf{MixGCF}~\cite{HDD21}, which leverages information from positive samples and graph neighborhood samples to synthesize negative samples based on unlabeled data.
    \item \textbf{ANS}~\cite{ZCL23}, which proposes to generate negative samples based on fine-granular factors within unlabeled data.
    \vspace{-2mm}
\end{itemize}

\subsection{RQ1: Overall Performance Comparison}
We report the main experimental results in Table~\ref{tab:Overall-comprehension-1} and Table~\ref{tab:Overall-comprehension-2}, where the reported improvements are the average of all metrics and are significant over the (best) baseline(s) under a two-sided t-test with $p<0.05$.
\begin{itemize}
    \item From Table~\ref{tab:Overall-comprehension-1}, it can be observed that, with the help of PNN, all base models show significant performance improvements in almost all cases. The performance improvement of NGCF on Yelp is as high as 30.07\%. Surprisingly, the experimental results indicate that, even with a simple BPR-MF model, PNN can achieve comparable performance to sophisticated GNN models. The results confirm the superiority and applicability of our proposed PNN solution.
    \item As shown in Table~\ref{tab:Overall-comprehension-2}, while traditional methods achieve reasonably good performance, PNN consistently delivers the best performance across all datasets and all metrics. The encouraging results suggest exploring a reasonable way to unearth and leverage the information within unlabeled data.
    \item It is interesting to observe that SRNS achieves better performance in some datasets. This validates our assertion that unlabeled data and negative samples exhibit an inevitable disparity. While SRNS improves performance by filtering items that users potentially like, unfortunately, due to its limited utilization of this filtered samples, its performance still lags behind PNN.
    \vspace{-2mm}
\end{itemize}

\begin{figure}[htpb]
\centering
\begin{minipage}[b]{0.50\linewidth}
    \centering
      \includegraphics[width=\linewidth]{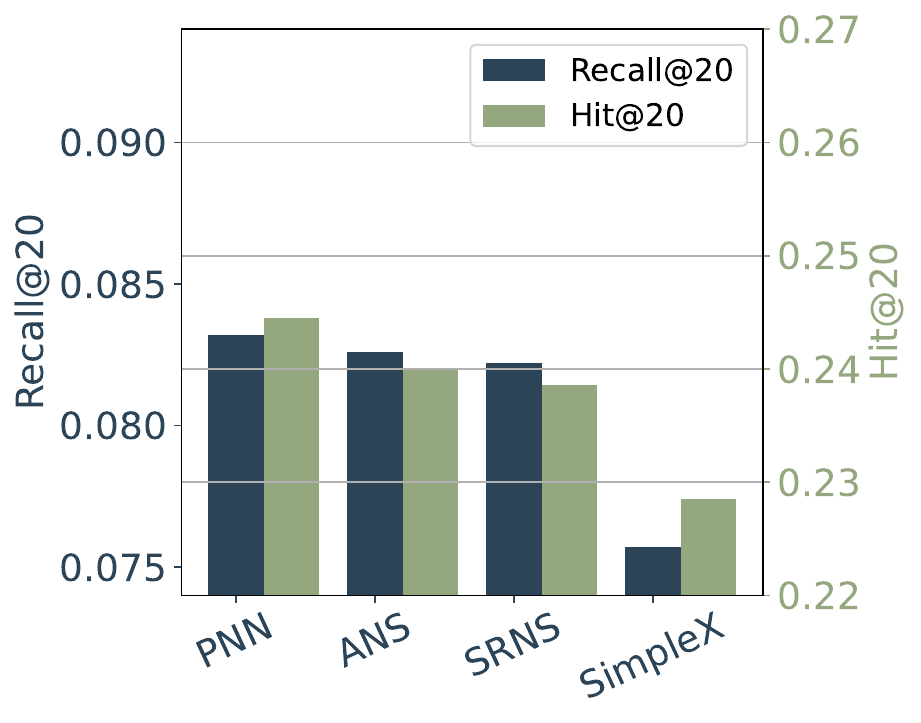}
      \subcaption{Yelp}
    \label{fig:Foursquare}
\end{minipage}
\hfill 
\begin{minipage}[b]{0.50\linewidth}
    \centering
    \includegraphics[width=\linewidth]{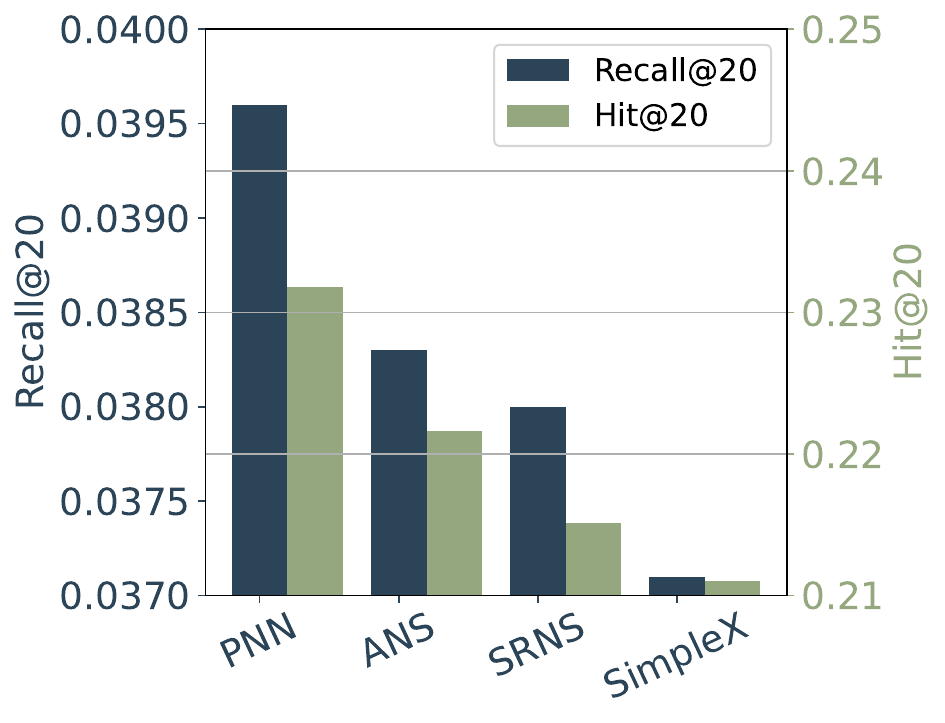}
    \subcaption{Foursquare}
    \label{fig:attribute_prediction}
\end{minipage}
\caption{The results of the study on neutral class.}
\label{fig:noise}
\vspace{-5mm}
\end{figure}

\subsection{RQ2: Study on Neutral Class}
In this section, we will explore the role of neutral samples in PNN. Intuitively, we can label items that users potentially like or items with user uncertainty in the dataset and then evaluate the detection rate of false negative samples. However, a high detection rate does not directly reflect the distinction between our method and traditional approaches (such as SRNS, which can also detect false negative samples). Inspired by~\cite{ZZH22}, we randomly select 10\% of interactions from Yelp and Foursquare datasets as negative to simulate the false negative samples and use the constructed datasets to evaluate different methods. Following the principles of PNN, these samples should be categorized as the neutral class and participate in training processes. As can be observed in Figure~\ref{fig:noise}, SimpleX performs worst because it relies on many negative samples, making it more vulnerable to noise. In contrast, PNN can effectively resist the impact of false negative noise. Despite SRNS's capability to filter false negative samples, its failure to effectively utilize this portion of the signal still results in suboptimal performance, reaffirming our motivation to fully leverage unlabeled data.

\subsection{RQ3: Ablation Study}
We analyze the effectiveness of different components via the following variants: (1) PNN without $\mathcal{L}_\mathrm{constrain}$ (PNN w/o constrain), (2) PNN without $\mathcal{L}_\mathrm{uniform}$ (PNN w/o uniform), (3) PNN without $\mathcal{L}_\mathrm{rank}$ (PNN w/o rank), (4) PNN without adaptive weight $\lambda$ (PNN w/o ada), and (5) PNN without semi-supervised learning (PNN w/o sem). The results are presented in Table~\ref{tab:ablation}. They suggest that all components positively contribute to model performance. We can observe that $\mathcal{L}_\mathrm{rank}$ (corresponding to PNN w/o rank) and semi-supervised learning (corresponding to PNN w/o sem) significantly impact the performance. The former can be attributed to the fact that the ranking task has a direct impact on the top-K recommendation task. Without $\mathcal{L}_\mathrm{rank}$, we cannot leverage the triple-wise ranking relationships. The latter confirms the challenge of the classification task in the absence of supervised signals and the effectiveness of the proposed semi-supervised learning method.

\begin{table}[t]
\caption{Ablation Study.}
\vspace{-2mm}
\begin{center}
\begin{tabular}{@{}lcccc@{}}
\toprule
\multirow{2}{*}{\textbf{Dataset}} & \multirow{2}{*}{\textbf{Method}}  &\multicolumn{3}{c}{Top-20}\\ 
\cmidrule(lr){3-5} & & Hit Ratio & Recall & NDCG  \\ 
\midrule
\multirow{3}{*}{\textbf{Yelp}} 
& PNN w/o constrain & 0.2705 & 0.0454 & 0.0389\\
& PNN w/o uniform & 0.2705 & 0.0451 & 0.0380 \\
& PNN w/o rank & 0.0960 & 0.0152 & 0.0115\\
& PNN w/o ada & 0.2761 & 0.0463 & 0.0399\\
& PNN w/o sem & 0.1967 & 0.0319 & 0.0262\\
& PNN  &\textbf{0.2770} &\textbf{0.0472} &\textbf{0.0405}\\
 \midrule

\multirow{3}{*}{\textbf{Gowalla}} 
& PNN w/o constrain & 0.3078 & 0.1562 & 0.0908\\
& PNN w/o uniform & 0.3140 & 0.1651 & 0.0951 \\
& PNN w/o rank & 0.0406 & 0.0150 & 0.0079\\
& PNN w/o ada & 0.3008 & 0.1558 & 0.0894\\
& PNN w/o sem & 0.1439 & 0.0620 & 0.0361\\
& PNN  &\textbf{0.3366} &\textbf{0.1787} &\textbf{0.1052}\\
\bottomrule
\end{tabular}
\label{tab:ablation}
\end{center}
\vspace{-4mm}
\end{table}

\subsection{RQ4: Hyperparameter Study}
\begin{figure}[htpb]
\centering
\begin{minipage}[b]{0.5\linewidth}
    \centering
      \includegraphics[width=\linewidth]{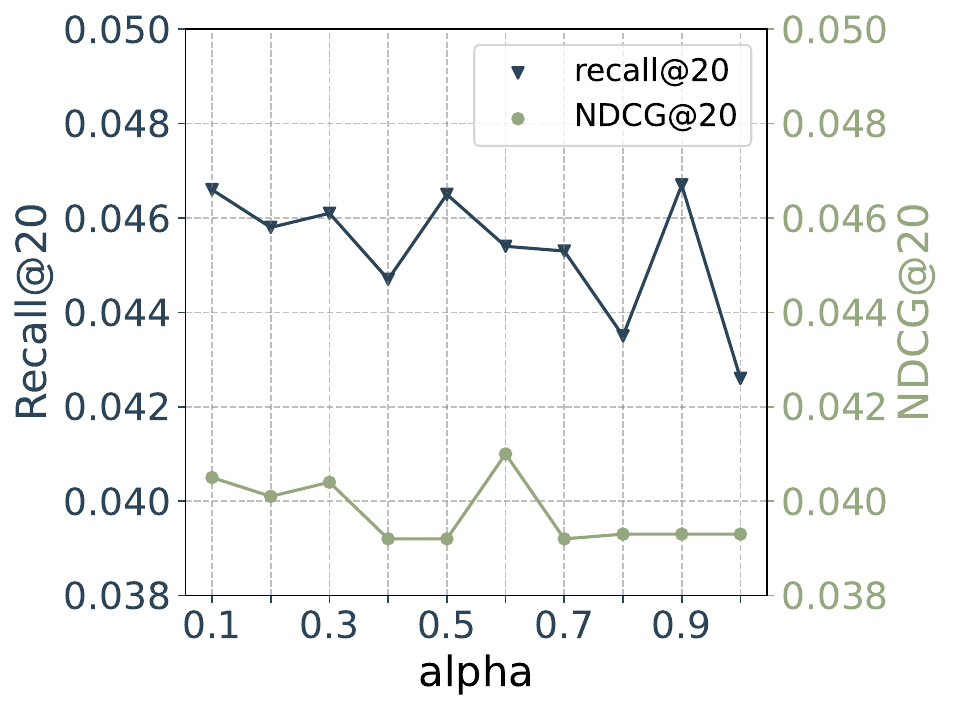}
      \subcaption{Foursquare}
    \label{fig:Foursquare}
\end{minipage}
\hfill 
\begin{minipage}[b]{0.5\linewidth}
    \centering
    \includegraphics[width=\linewidth]{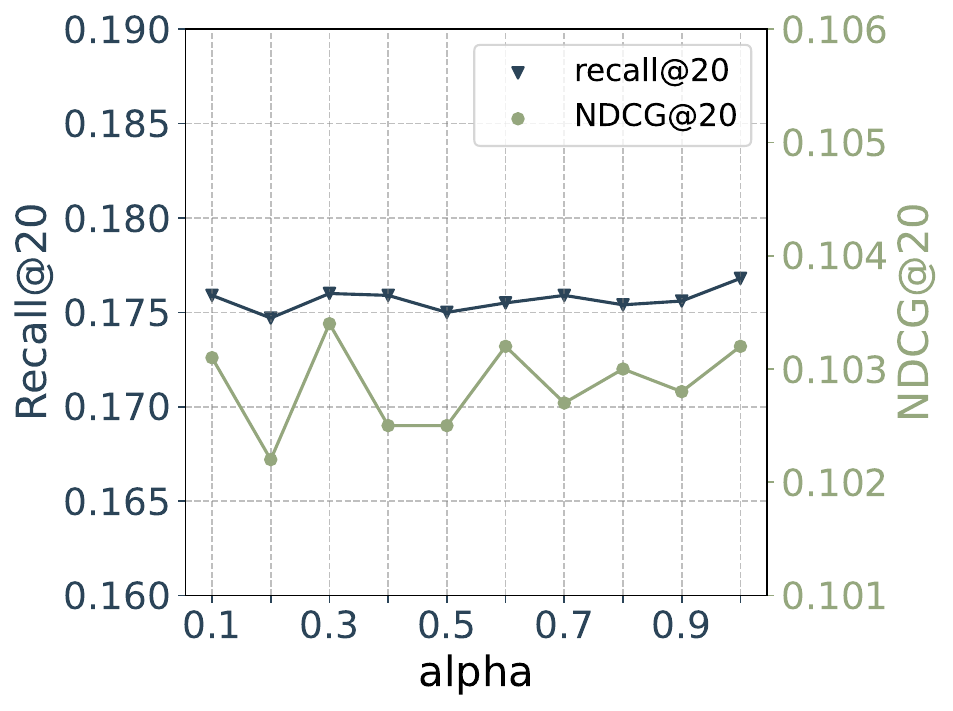}
    \subcaption{Gowalla}
    \label{fig:attribute_prediction}
\end{minipage}

\begin{minipage}[b]{0.5\linewidth}
    \centering
      \includegraphics[width=\linewidth]{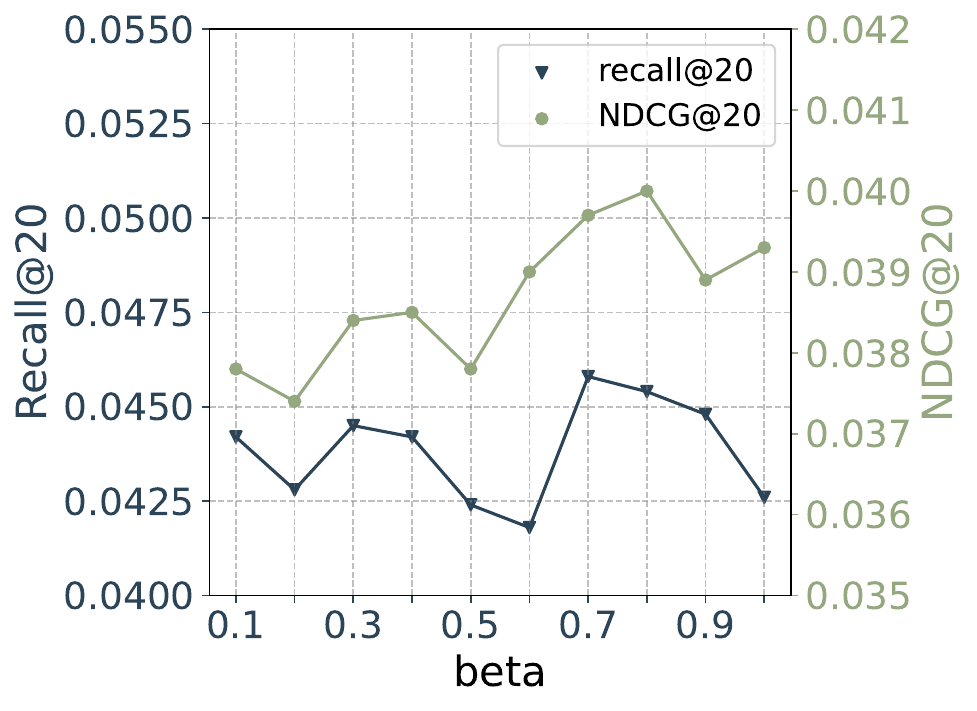}
      \subcaption{Foursquare}
    \label{fig:Foursquare}
\end{minipage}
\hfill 
\begin{minipage}[b]{0.5\linewidth}
    \centering
    \includegraphics[width=\linewidth]{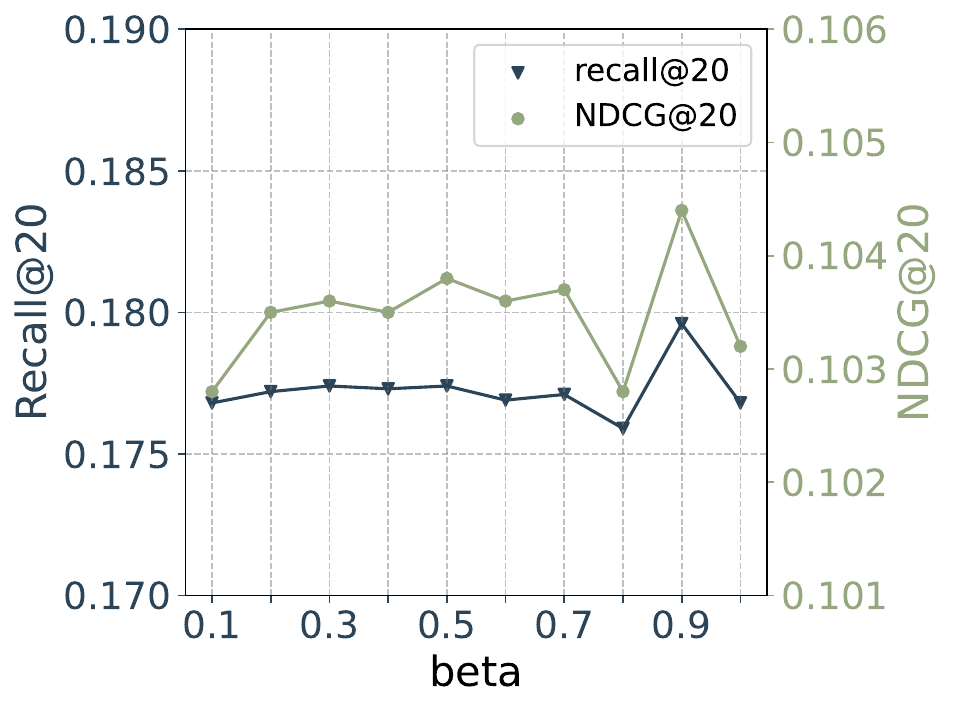}
    \subcaption{Gowalla}
    \label{fig:attribute_prediction}
\end{minipage}
\caption{Impact of hyperparameters on PNN.}
\label{fig:H}
\vspace{-3mm}
\end{figure}
We conduct experiments on Foursquare and Yelp to study the impact of different values of $\alpha$ and $\beta$ and present the results in Figure~\ref{fig:H}. Recall that $\alpha$ is the parameter to balance the importance of $\mathcal{L}_\mathrm{constrain}$ and $\beta$ to balance the importance of $\mathcal{L}_\mathrm{uniform}$. An overly small $\alpha$ value may lead to class overlap and cannot preserve the desirable set-level ranking property; an overly large $\alpha$ value may create large margins that push intra-class items to cluster. Similarly, overly large or small $\beta$ values could undermine model performance. The experimental results are well aligned with our analysis. But overall PNN can obtain good performance under a relatively wide range of hyperparameters. 



\begin{table}[t]
\caption{The efficiency results of different methods.}
\begin{center}
\vspace{-2mm}
\begin{tabular}{@{}lcccc@{}}
\toprule
\multirow{1}{*}{\textbf{Dataset}} & \multirow{1}{*}{\textbf{Method}}   & \multirow{1}{*}{\textbf{Training Time}} & \multirow{1}{*}{\textbf{\# of Epoch}} \\ 
\midrule
\multirow{3}{*}{\textbf{Yelp}} 
& BPR-MF & 2.40s & 82 \\
& SimpleX & 431.92s & 36 \\
& ANS & 258.26s & 16  \\
& PNN & 250.46s & 15 \\

 \midrule
\multirow{3}{*}{\textbf{Gowalla}} 
& BPR-MF & 4.29s & 47 \\
& SimpleX & 342.33s & 29 \\
& ANS & 194.63s & 20  \\
& PNN & 45.66s & 27 \\

\bottomrule
\end{tabular}
\label{tab:Efficiency}
\end{center}
\vspace{-4mm}
\end{table}

\subsection{RQ5: Efficiency Analysis}
Following the previous works~\cite{MZW21,WYM22,ZCL23}, we compare the efficiency of PNN with other representative methods (BPR-MF, SimpleX, and ANS). Table~\ref{tab:Efficiency} presents the average training time per epoch and the number of epochs required to converge. All methods are relatively efficient. As expected, BPR-MF is the most efficient method due to its simplicity. SimpleX demonstrates the worst efficiency due to its requirement of a large number of negative samples and the subsequent filtering of uninformative negative samples. PNN is the second most efficient method. With its outstanding performance, we believe it is a desirable choice for practical use.

%% file: 2.related_work.tex
\section{RELATED WORK}
\label{related_work}
Effectively leveraging the massive unlabeled data is a significant challenge. Most existing works~\cite{RFG12,CZW19,ZCW13,DQH19,CZZ20} make an intuitive assumption that unlabeled data can directly provide negative signals. BPR loss function~\cite{RFG12, SGZ20,YYG21} randomly selects unlabeled data as negative samples. ENMF~\cite{CZZ20} efficiently learns model parameters from the whole unlabeled data. Yet, recent works reveal a nuanced reality: unlabeled data and negative samples harbor an inevitable disparity. Acknowledging this phenomenon, current research attempts to extract a subset closely approximating negative samples. SimpleX~\cite{MZW21} extracts a large volume of high-information samples from unlabeled data to serve as negative samples. Zhuo \textit{et al.}~\cite{ZZY22} introduces the concept of user interest boundary (UIB) to penalize samples crossing a threshold in the ranking task. DNS~\cite{ZCW13} selects unlabeled data that are more similar to a user as negative samples. GDNS~\cite{ZZH22} develops a gain-aware function to select real negative samples. SRNS~\cite{DQY20} selects negative samples while incorporating a variance mechanism to filter out false negative samples. MixGCF~\cite{HDD21} integrates information from a graph structure and positive samples to enhance the hardness of negative items. DENS~\cite{LCZ23} disentangles relevant and irrelevant factors of samples to select appropriate negative samples. ANS~\cite{ZCL23} proposes to generate synthetic negative samples to improve implicit CF. Regrettably, under this approach, the remaining data are overlooked, we want to unlock these hidden treasures behind unlabeled data to improve recommendation performance.

%% file: 7.conclusion.tex
\section{CONCLUSION}
\label{conclusion}
In this work, our paper introduces an innovative positive-neutral-negative (PNN) learning paradigm, which changes the utilization of massive unlabeled data in CF for recommender systems. By introducing a third neutral class and leveraging set-level ranking relationships, PNN effectively addresses the complexity inherent in unlabeled data. We have provided a concrete implementation of PNN that seamlessly integrates with multiple mainstream CF models. This implementation includes a novel semi-supervised learning method with a user-aware attention model to classify unlabeled data reliably, as well as a two-step centroid ranking approach to handle set-level rankings. Through extensive experiments on four public datasets, we demonstrate that PNN significantly improves the performance of various mainstream CF recommendation models. Our work marks a significant advancement in recommender systems, promising improved user experiences and broader applicability in real-world scenarios.

%% file: acks.tex
\begin{acks}
This work was supported by the Heilongjiang Key R\&D Program of China under Grant No. GA23A915, the National Natural Science Foundation of China under Grant No. 62072136, and partially supported by the Hong Kong Baptist University IG-FNRA project (RC-FNRA-IG/21-22/SCI/01).
\end{acks}